\documentclass[aps,pra,groupedaddress,superscriptaddress,twocolumn]{revtex4-1}

\usepackage{graphicx}  
\usepackage{amssymb}   
\usepackage{amsmath}
\usepackage[usenames,dvipsnames]{color}
\usepackage[colorlinks=true, citecolor=Red,urlcolor=blue, linkcolor=BrickRed]{hyperref}

\renewcommand{\emph}{\textit}

\begin{document}

\title{Exploring Localization in Nuclear Spin Chains}

\author{Ken Xuan Wei}
\affiliation{Department of Physics \& Research Laboratory of Electronics,
Massachusetts Institute of Technology, Cambridge, MA 02139, USA}
\author{Chandrasekhar Ramanathan}
\affiliation{Department of Physics and Astronomy, Dartmouth College, Hanover, NH 03755, USA}
\author{Paola Cappellaro}
\affiliation{Department of Nuclear Science and Engineering \& Research Laboratory of Electronics,
Massachusetts Institute of Technology, Cambridge, MA 02139, USA}
\email[]{pcappell@mit.edu}

\date{\today}

\begin{abstract}
Characterizing out-of-equilibrium many-body dynamics is a complex but  crucial task for quantum applications and the understanding of fundamental phenomena. A central question is the role of localization  in quenching  quantum thermalization, and  whether  localization   survives in the presence of interactions. 
The localized phase of interacting systems (many-body localization, MBL) exhibits a long-time logarithmic growth in entanglement entropy that distinguishes it from the noninteracting Anderson  localization (AL), but entanglement is difficult to measure experimentally.  
Here, we present a novel correlation metric, capable of distinguishing MBL from AL in high-temperature spin systems. 
We demonstrate the use of this metric to detect localization in a natural solid-state spin system using nuclear magnetic resonance (NMR). We  engineer the natural Hamiltonian to  controllably introduce disorder and interactions  and   observe the emergence of localization. 
In particular, while our correlation metric saturates for AL,  it keeps increasing logarithmically for MBL,  a behavior reminiscent of entanglement entropy, as we confirm by simulations.  Our results show that our NMR techniques, akin to measuring out-of-time correlations, are well suited for studying localization in spin systems.
\end{abstract}

\maketitle

Anderson first demonstrated that single particle wave functions can become exponentially localized in the presence of disorder~\cite{Anderson58}. Whether this localization~\cite{Wiersma97, Billy08, Roati08} survives in the presence of interaction has received much attention in recent years~\cite{Basko06,Basko07,Nandkishore15,Smith16,Schreiber15,Choi16}. Numerical evidence in spin chains indicates that the system may be in the MBL or ergodic phase depending on the relative strength of interaction and disorder~\cite{Pal10,Serbyn15,Luitz15}. Furthermore, the MBL phase is distinct from its noninteracting counterpart, AL, in the dynamics of entanglement entropy~\cite{Bardarson12,Serbyn13,Huse14}. 
Entanglement entropy is however difficult to evaluate experimentally, and so far has only been measured on systems with small number of particles~\cite{Islam15}. One way to circumvent this is to measure entanglement witnesses such as the quantum fisher information (QFI), which can serve as a lower bound for entanglement entropy~\cite{Smith16} for pure states.

The MBL phase is predicted to persist at high and even  infinite temperature~\cite{Oganesyan07}, where states are highly mixed and there is little to no entanglement present. How does one characterize the MBL phase experimentally in such a system? 
Here we introduce a novel metric capable of distinguishing MBL from AL in non-equilibrium dynamics of highly mixed states. 
Our approach requires no local control of the system, and relies only on collective rotations and measurements, in contrast to recently proposed metrics~\cite{Goihl16} that also detect the spread of correlations but require single-spin addressability. 
We provide numerical and experimental evidences of this metric. In particular, we report on experimental observations of AL and MBL by measuring the evolution of many-spin correlations.  The experimental system is composed of nuclear spins in a natural crystal coupled by the magnetic dipolar interaction. While the system is a 3D, open quantum system with long-range interactions, it has been shown~\cite{Zhang09,Cappellaro07l} that for timescale of relevance in the experiments it can mapped with high-fidelity to an ensemble of 1D, nearest-neighbor coupled spin chains. Here we further show that we can exploit Hamiltonian engineering and other spins in the system to introduce  tunable interaction and disorder. 
\begin{figure}\centering
\includegraphics[width=0.35\textwidth]{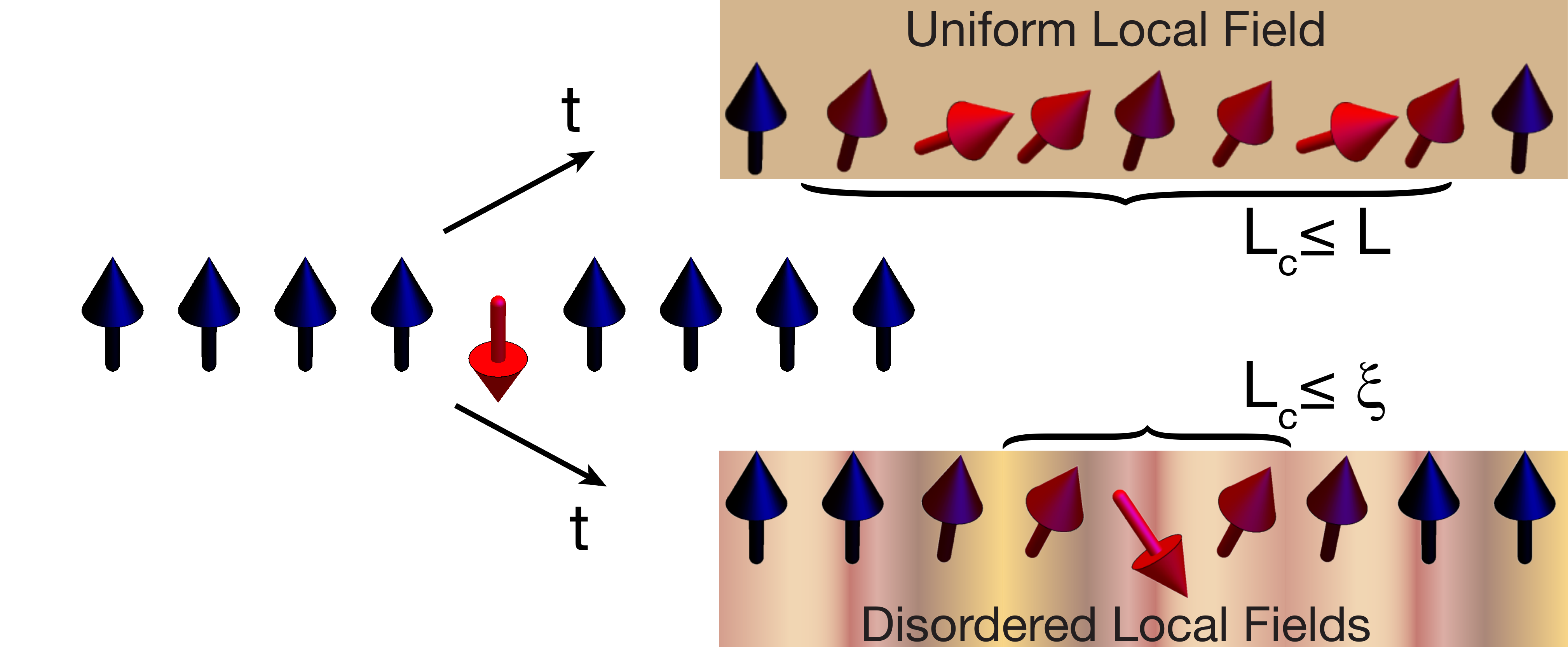}
	\caption{Quantum many-body correlations (top) grow from an initial localized state (left) but are restricted to a finite size by disorder (bottom). The average correlation length $L_c$ measure the spread of the correlations.}
\label{fig:model}
\end{figure}

\begin{figure*}[t]
\centering    
\includegraphics[width=0.8\textwidth]{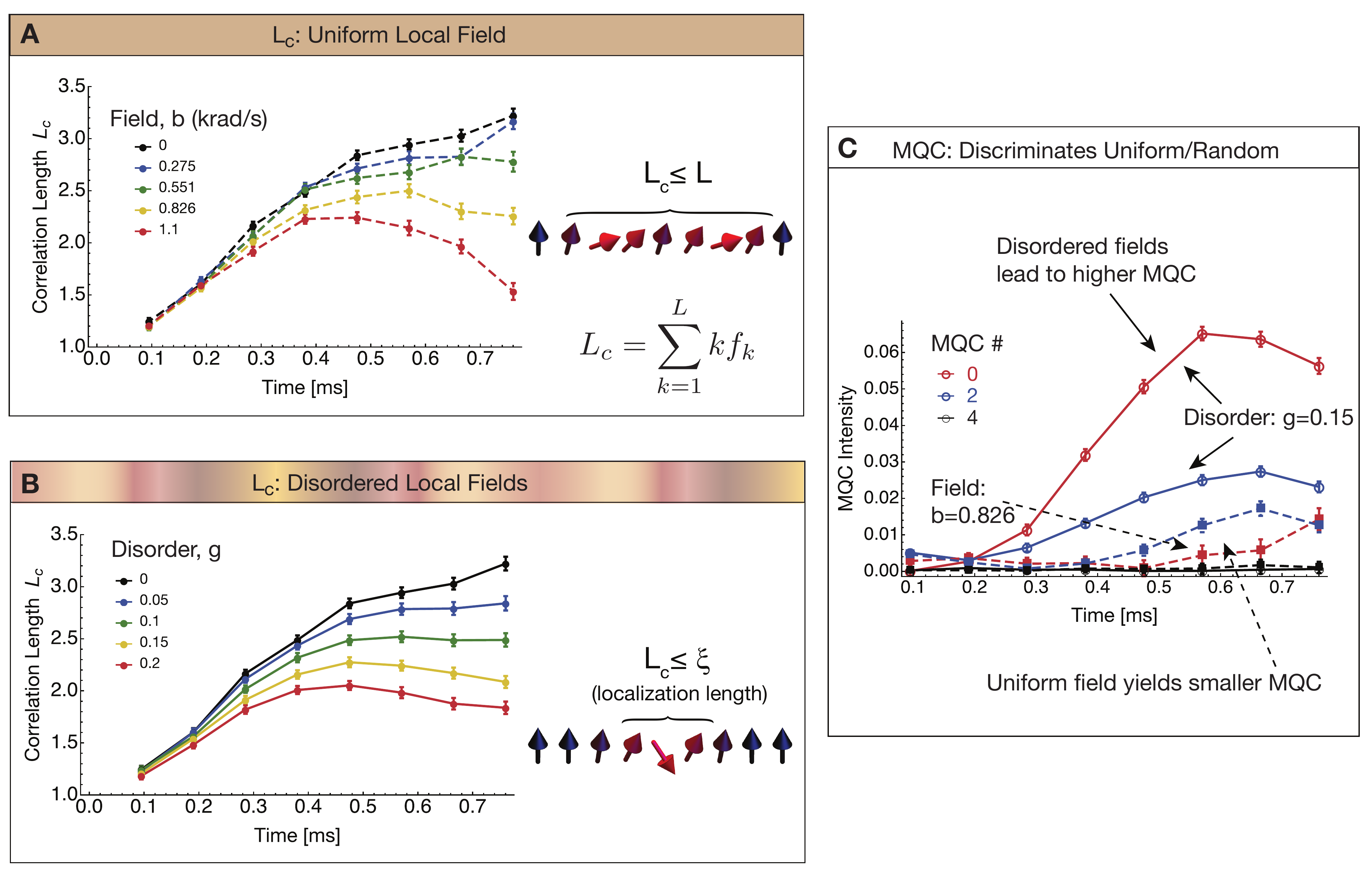}
\caption{
\textbf{Experimental measurements of spin correlations in noninteracting spin chains.} Correlation length $L_c$ for uniform (\textbf{A}) and disordered transverse fields (\textbf{B}). In both cases we set $u=0.2$ and $v=0$ and varied the disorder strength $g$ and field magnitude $b$ (see Eq.~\ref{eq:Hamiltonian}). (\textbf{C}) Comparison of MQC intensities in the $\chi$ sector used  as a litmus test for disorder, see also SM. Solid markers are for uniform field  with $b=0.826$ krad/s, open markers are for disordered field  with $g=0.15$. Errorbars are determined from the noise in the free induction decay, the solid lines are guides to the eye.}
\label{fig:AL}
\end{figure*}
%
Our experimental system consists of a single crystal of fluorapatite [Ca\(_5\)(PO\(_4\))\(_3\)F] placed in a strong magnetic field (7 Tesla) along \(z\). 
The \(^{19}\)F spin-1/2 nuclei in the hexagonal fluorapatite crystal form linear chains along the \(c\)-axis, each surrounded by six other chains.  When the \(c\)-axis is oriented parallel to the external magnetic field, the cross-chain coupling is 40 times weaker than intra-chain coupling. The system can be treated approximately as an ensemble of identical spin chains~\cite{Cappellaro07l,Cappellaro11,Ramanathan11}. In addition, each F spin is surrounded by three \(^{31}\)P spin-1/2 nuclei.

The spins interact via the natural dipolar Hamiltonian, \(H_\textrm{nat}=\frac{1}{2}\sum_{j<k}J_{jk}(2\sigma_z^j \sigma_z^{k}-\sigma_x^j \sigma_x^{k}-\sigma_y^j \sigma_y^{k})+\sum_{j,\kappa} h_{j\kappa}\sigma_z^j s_z^\kappa\), where \(\sigma_\alpha^j\) \((\alpha=x,y,z)\) are  Pauli matrices of the \(j\)-th F spin and \(s_z^\kappa\)  of the \(\kappa\)-th P spin. The  two terms  are, respectively, the homonuclear dipolar interaction between F spins and the heteronuclear dipolar interaction between F and P spins. 
At room temperature 
the P spins are in an equal mixture of \(m_z=\pm1/2\) states. This allows us to replace the heteronuclear interactions by \(\sum_j h_j \sigma_z^j\), where \(h_j\) is now a random variable representing the disordered field seen locally by each fluorine.  

Even if the natural Hamiltonian does not directly lend itself to study  localization, we can perform a sudden quench to the desired effective Hamiltonian  by periodically applying a radiofrequency pulse sequence in resonance with the F spins.  This method (called average Hamiltonian theory~\cite{Haeberlen76}) has been long used in the NMR literature for spectroscopy and condensed matter studies. 
Here we further push these techniques  to engineer a broad class of Floquet (periodic) Hamiltonians with tunable disorder and interactions. In addition, we are also able to reverse the arrow of time, a tool that allows measuring  out-of-time ordered correlations (OTOC).  
As shown in the supplementary material (SM), the effective Hamiltonian can be written as 
\begin{align}
\label{eq:Hamiltonian}
H=&u  \sum_{j=1}^{L-1}\frac J2(\sigma_x^j \sigma_x^{j+1}-\sigma_y^j \sigma_y^{j+1})+b\sum_{j=1}^{L} \sigma_z^j  
\\+&g \sum_{j=1}^{L} h_j \sigma_z^j+ v  \sum_{j=1}^{L-1}\frac J2(\sigma_x^j \sigma_x^{j+1}+\sigma_y^j \sigma_y^{j+1} - 2\sigma_z^j  \sigma_z^{j+1})\nonumber ,
\end{align}
with \(J\)  the nearest-neighbor coupling, \(b\) a uniform  field, and \(h_j\) the disordered field provided by background P spins, as well as other nuclear and electronic spin defects in the system.  
The first two terms represent an integrable Hamiltonian, as they map to a free fermionic Hamiltonian. The last two terms introduce disorder and interaction, respectively. In particular, the term $\sigma_z^j\sigma_z^{j+1}$ maps into fermion density-density interactions (see Eq. (2) in SM).
The experimentally adjustable parameters  \(u\), \(v\), \(g\), and \(b\)  allow us to explore various regimes of interest.

We consider a linear chain of $L$ spins initially at equilibrium at high temperature, $\beta\approx0$. To first order in $\epsilon =\beta\omega_L/2$, its state can be expressed as \(\rho_\textrm{eq}=(\openone-\epsilon \sum_j \sigma_z^j)/2^L\) ($\omega_L$ is the spin Zeeman energy and we set $\hbar=1$). 
When an Hamiltonian \(H\) is applied with a sudden quench, the system evolves  into a many-body correlated state. 
Disorder hinders the growth of correlations and gives rise to localized states,  characterized by an exponentially decreasing probability of  correlations outside a typical localization length, $\xi$.
Inspired by this  picture, we define a metric of localization that measures the average length over which correlations have developed. 
More precisely, we can generically write  the high-temperature time-evolved density matrix  as
\begin{align}
\rho(t)=\frac\openone{2^L}-\frac{\epsilon \sqrt{L}}{2^L} \sum_{k=1}^L \sum_{s=1}^{\zeta_k}b^{s}_k(t)\mathcal{B}^{s}_{k},
\label{eq:rhoN}
\end{align}
where \(\mathcal{B}^s_k\) 
are operators composed of tensor products of $k$ Pauli matrices and $L-k$ identity operators. 
To quantify localization  we then define 
 the \textbf{\textit{average correlation length}}
\begin{align}
L_c=\sum_{k=1}^{L} k f_k,
\label{eq:Nc}
\end{align}
where 
\(f_k=\sum_{s=1}^{\zeta_k}[b_k^s]^2\) is the contribution of all possible spin correlations with Hamming weight $k$ (with \(\sum_{k=1}^L f_k =1\)). In the SM, we introduce a closely related metric, the \textit{average correlation distance}, $D_c$, that can be analogously used to distinguish between AL and MBL, but  is more difficult to measure experimentally.

In the initial equilibrium state \(\rho_\textrm{eq}\)  there are  no spin correlations and \(L_c=1\). 
In a non-disordered system, we expect \(L_c\) to grow indefinitely (see SM) or, for a finite system, to  eventually saturate at a value dependent on \(L\).
Introducing disorder leads to a quantitatively different behavior. 
When the system is noninteracting, AL leads to a coherent suppression of many-spin correlations and \(L_c\) is bound by { the localization length} \(\xi\). 
When interactions are present, disorder is unable to completely suppress the correlation  growth. The slow growth of \(L_c\) in the presence of interactions is the key feature that enables \(L_c\) to distinguish between AL and MBL for mixed states.

Determining \(f_k\) for a generic many-body state is  challenging, since it is usually difficult to directly measure many-body correlations and  the number of configurations \(\zeta_k\) is exponential in \(k\) and \(L\). 
Since our metric is based on the number of correlated spins, we can however  borrow from well-known NMR techniques that approximate the number of correlated spins by their quantum coherence number~\cite{Munowitz75}. 
Multiple quantum coherence (MQC) intensities of order \(q\) describe the contribution of terms \(|m_a\rangle\!\langle m_a'|\) in the density matrix such that \(m_a-m_a'=q\), with \(m_a\) the collective \(\sigma_a\) eigenvalue (typically $a=z$).  MQC intensities \(I_q\) can be measured by relying on their distinct behavior under collective rotations~\cite{Munowitz75,Baum85}. 
The distribution of \(I_q\) has been traditionally used to approximate the average number of correlated spins, or cluster size, in 3-D spin networks~\cite{Baum86,Munowitz87,Alvarez15}. While this approximation fails in 1-D systems, we  find instead a practical experimental protocol to \textit{exactly} measure \(L_c\) for noninteracting systems. The protocol still yields a very good approximation for disordered interacting (MBL) systems. 

We first note that in noninteracting systems, for simple initial states such as \(\rho_\textrm{eq}\) the number of configurations is simply \(\zeta_{k}\propto L-k\): All many-spin correlations are in the form \(\mathcal B^s_k\sim\sigma_a^s (\prod_{l=s+1}^{k+s-2}\sigma_z^l) \sigma_b^{k+s-1}\), where the end spins \(\sigma_{a,b}\) are either \(\sigma_x\) or \(\sigma_y\). 
Then, correlations with different \(k\) will  respond differently when rotated around an appropriate axis. 
In our MQC protocol, we first decompose \(\rho(t)\) into four orthogonal sectors using time-reversal and phase cycling~\cite{Bodenhausen84}  and then measure the MQC intensities encoded in the \(x\) axis for each $j^{\textrm{th}}$ sector, \(I^j_q\) (see SM). 
The resulting MQC intensities can be related to \(f_k\) in Eq.~(\ref{eq:Nc}) by a linear transformation, \(f_k\!=\!\sum_{jk} M^{(j)}_{kq}I^j_q\), and from the extracted $f_k$  we can   calculate \(L_c\). 

\begin{figure*}[tb]
\centering 
\includegraphics[width=0.95\textwidth ]{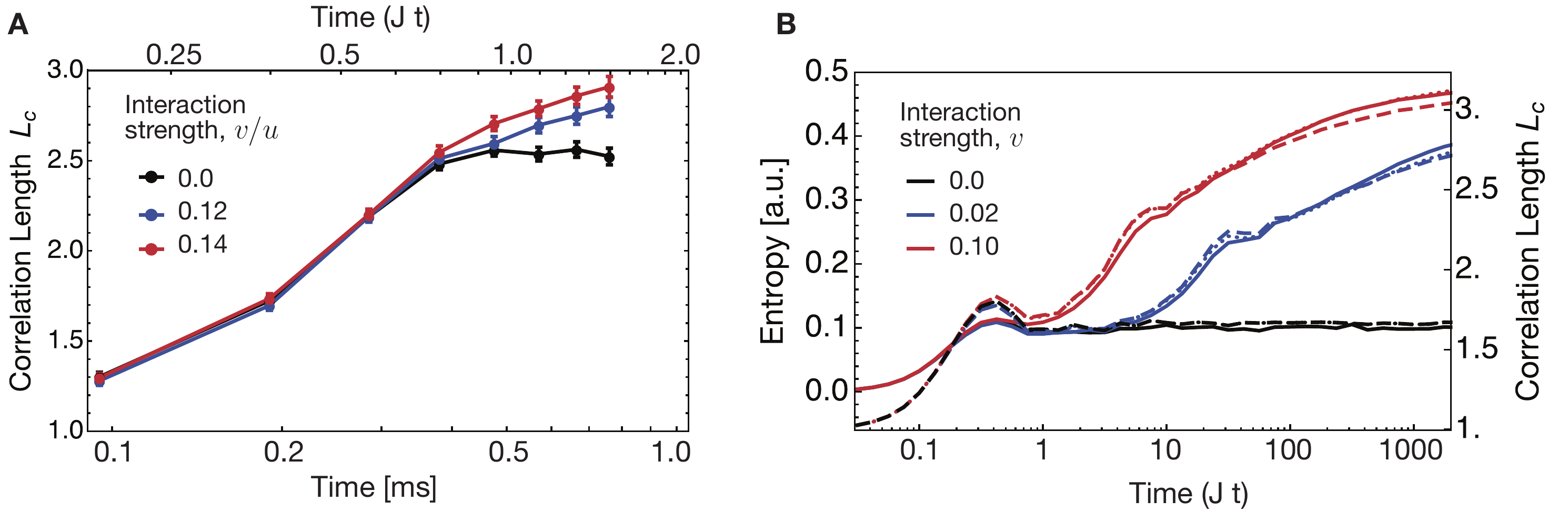}\caption {\textbf{A: Experimental measurements of spin correlations in interacting spin chains.} We plot in log-linear scale the measured \(L_c\) dynamics in the presence of disorder and  for varying interaction strengths $v$. Data are for \(u=0.24\), \(g=0.12\), and \(b=0\). After an initial growth of correlations, $L_c$ saturates for the non-interacting systems, while it shows a slow growth in the presence of interactions, thus indicating many-body localization.
\textbf{B: Simulations of spin correlation and entanglement entropy.} We compare the entropy of the reduced half chain (solid lines, left axis) with the correlation length $L_c$ (dotted lines, right axis) and the approximate $L_c$ obtained from measuring the MQC (dashed lines). The similar behavior (including logarithmic growth) confirm that the chosen metric is as good an indicator of MBL as the more commonly used entanglement entropy. Here we renormalized the entanglement entropy to vary between 0 and 1, see SM for details.
}
\label{fig:MBLlog}
\end{figure*}

We first use the noninteracting version of our model (\(v=0\)) to show that we can indeed  measure the correlation length, and that its saturation is due to the controlled insertion of  disorder  and not simply to experimental imperfections. To that end, we use properties of the MQC intensities to discriminate between disorder and a uniform transverse field (Fig.~\ref{fig:AL}). 
In the absence of disorder we expect \(L_c\) to increase linearly, consistent with the Lieb-Robinson bound for short-ranged Hamiltonians~\cite{Lieb72}. 
In the thermodynamic limit \(L\rightarrow \infty\) and at large times \(uJt\gg 1\), \(L_c\) grows with a velocity \(V=8u J/\pi\). 
When a transverse field is present, \(L_c\) grows with a reduced velocity \(V\approx\frac{2 J^4 t u^4}{\pi  \left| b\right| ^3}\) and an oscillatory behavior (see SM). In both cases these is no saturation, except for finite size effects.
In the presence of disorder, instead, we expect \(L_c\) to initially increase, as spins correlate within the localization length, and to saturate at long times due to AL. 
In the experiments we can probe this dynamics only for relatively short times, where the physical system is a good approximation to the ideal model~\cite{Zhang09}. Indeed, we have a 3-D crystal where each spin chain interacts with 6 surrounding chains and the couplings are long-range, $\propto1/r^3$.  
Thus, we kept the experimental time short to minimize these effects, as well as pulse imperfections that can lead to unwanted terms in the engineered Hamiltonian (the time is also much shorter than the relaxation time $T_1\approx 0.8$~s and the P dynamics). 
On these timescales, however, \(L_c\) shows an apparent saturating behavior for both disordered or transverse fields. 
In order to claim AL in the  disordered case, we show that there are distinctive features in the MQC spectrum (see SM): 
An additional symmetry in the transverse field case forces  one of the MQC sectors to vanish in the absence of disorder, while it is still non-zero for  the disordered case, as shown in Fig.~\ref{fig:AL}.\textbf{C}. 
This indicates that the behavior of \(L_c\) in Fig.~\ref{fig:AL}.\textbf{A} is not due to disorder, and we expect \(L_c\) to resume linear growth at long times. Conversely, we can use this experimental evidence to prove that our Hamiltonian engineering technique can indeed introduce disorder in the system evolution.

We next study the behavior under interactions by varying the value of $v$ in Eq.~(\ref{eq:Hamiltonian}).
Fig.~\ref{fig:MBLlog}.\textbf{A}  shows the experimentally extracted \(L_c\) for our interacting model, as compared to the non-interacting case. The experiments clearly reveal the emergence of slow growth in \(L_c\) when interactions are added, the hallmark feature of MBL~\cite{Bardarson12,Pino14}. 
To confirm that indeed we can observe MBL with the correlation length metric, we compare  its behavior to the well-known dynamics of the entanglement entropy in numerical simulations.

We first check that the interacting version of our model (\(v\neq 0\)) indeed supports MBL by calculating the von Neumann entropy for an initial pure state evolving under the Hamiltonian in Eq.~(\ref{eq:Hamiltonian}).
The  entropy  \(S=-\textrm{Tr}[\rho_L\log\rho_L]\), where \(\rho_L\) is the reduced density matrix of the left half of the chain, displays a characteristic logarithmic growth in time~\cite{Bardarson12} when the system enters the MBL phase (see SM).
Next, for the initial equilibrium state $\rho_\textrm{eq}$, 
we compare in Fig.~\ref{fig:MBLlog}.B the simulated \(L_c\)  with the entropy of the reduced half-chain state (approximated to second order in $\epsilon$, see Eq.~\ref{eq:rhoN}). 
We find that both $L_c$ and $S$ saturate at long times when the system is noninteracting and increases logarithmically when a weak interaction is introduced. 
This suggests that $L_c$  can  be used  as a measure of  entanglement entropy to distinguish MBL from AL for mixed states. Finally, we check that our experimental method for measuring \(L_c\) via its relation to the  MQC is still a good approximation even when introducing interactions. 
We find that the approximated \(L_c\) also shows logarithmic growth in the MBL phase, 
with the approximation becoming better with increasing disorder, i.e., in the deep MBL phase.

Besides these numerical evidence, we can further obtain a more intuitive understanding of why our experimental method for extracting the correlation length from MQC works in a quite robust way. 
As mentioned, in the case of a non-interacting Hamiltonian, only a restricted set of operators \(\mathcal B_{k}^r\) appear in the dynamics and we can exactly measure their contribution to \(L_c\).  
For a Hamiltonian in the  MBL phase, the interactions are only a perturbation and thus we still expect a similar behavior. What is more, while in principle the number of possible configurations \(\zeta_k\) that could be populated is  exponential, in the presence of disorder only a fraction of them  have significant weights (following the area law~\cite{Eisert10}). Then, when applied to MBL systems, the MQC method effectively  undercounts the true \(L_c\), but still exhibits the same logarithmic growth. 
We can further understand our measurement in terms of out-of-time ordered correlations~\cite{Huang16,Garttner16x,Li16x}. As explained in details in the SM, 
to extract the MQC intensities we effectively measure the quantities
\begin{align}
\label{eq:otoc}
  S_\phi(t)=&\textrm{Tr}\left[\rho_\textrm{eq}\Phi^\dag(t)\rho_\textrm{eq}\Phi(t)\right],\\
\textrm{with}\quad  &\Phi(t)=U(t)e^{i\frac{\phi}{2} \sum_j\sigma_x^j}U^\dag(t).\nonumber
\end{align}
 While we can only measure OTOC for collective operators on the whole system, such as $\Phi$, these OTOC still give some information about the spreading or localization of correlations, since $\rho_\textrm{eq}$ is a sum of local operators. The information is made more accurate as we consider an average of several OTOC for different $\Phi(0)$ operators, even if we cannot measure a whole basis of a subsystem as required to extract the entropy~\cite{Hosur16,Huang16,Fan16x}.
It will be interesting to experimentally measure other OTOC  in our system, as OTOC has been studied in the context of information scrambling in black holes~\cite{Swingle16,Yao16x}.

In conclusion, we introduced a novel metric for localization, able to distinguish between many-body and single-particle localization. The correlation metric can be measured experimentally, with the only requirement of collective rotations and measurements, by extending MQC techniques developed in NMR that can as well be applied to many other physical systems~\cite{Garttner16x}. We further revealed an interesting relationship between the protocol for measuring the correlation length and the measurement of OTOC, thus further confirming its ability to measure the logarithmic growth of entanglement associated with MBL. 
Thanks to our control techniques, we were able to explore a broad range of interesting behaviors in our solid-state spin system. In particular, we  observed for the first time many-body localization in a natural spin system associated with a single crystal at high temperature. We note that while we interpreted our results mostly based on a simplified model (1D, nearest-neighbour couplings), the real system is more complex due to long-range interactions and a 3D structure. It will be thus interesting to use the tools developed in this work to study subtler properties of localization when these effects are highlighted by the experimental scheme. 

\begin{acknowledgements}
It is a pleasure to thank SoonWon Choi, Iman Marvian,  Seth Lloyd, and   Mikhail Lukin for insightful discussions.
This work was supported in part by the U.S. Air Force Office of Scientific Research grant No. FA9550-12-1-0292,  the U.S. Office of Naval Research grant No. N00014-14-1-0804, and by the National Science Foundation PHY0551153.
The authors declare that they have no competing interests.   
\end{acknowledgements}

\bibliographystyle{apsrev4-1}
\bibliography{Biblio}

\newpage
\onecolumngrid
\newpage

\section*{SUPPLEMENTARY MATERIAL}
\appendix

\section{Experimental System}
The system used in the experiment was a single crystal of fluorapatite (FAp).  Fluorapatite is a hexagonal mineral with space group \(P6_3/m\), with the \(^{19}\)F spin-1/2 nuclei  forming linear chains along the \(c\)-axis. Each fluorine spin in the chain is surrounded by three \(^{31}\)P spin-1/2 nuclei.
We used a natural crystal, from which we cut a sample of approximate dimensions 3 mm$\times$3 mm$\times$2 mm.
The sample is placed at room temperature inside an NMR superconducting magnet producing a uniform $B=7$ T field. The total Hamiltonian of the system is given by
\begin{equation}
H_{tot}=\frac{\omega_F}{2}\sum_k\sigma_z^k+\frac{\omega_P}{2}\sum_\kappa s_z^\kappa+H_{F}+H_P+H_{FP}
\label{eq:Hamtot}	
\end{equation}
The first two terms represent the Zeeman interactions of the F($\sigma$) and P($s$) spins, respectively, with frequencies $\omega_F=\gamma_FB\approx (2\pi)282.37$ MHz and $\omega_P=\gamma_PB=(2\pi)121.51$ MHz, where $\gamma_{F/P}$ are the gyromagnetic ratios. The other three terms represent the natural magnetic dipole-dipole interaction among the spins, given generally by
\begin{equation}
  H_{dip}=\sum_{j<k}\frac{\hbar\gamma_j\gamma_k}{4|\vec r_{jk}|^3}\left[\vec \sigma_j\cdot\vec\sigma_k-\frac{3\vec\sigma_j\cdot\vec r_{jk}\,\vec\sigma_k\cdot\vec r_{jk}}{|\vec r_{jk}|^2}\right],
\end{equation}
where $\vec r_{ij}$ is the vector between the $ij$ spin pair. Because of the much larger Zeeman interaction, we can truncate the dipolar Hamiltonian to its energy-conserving part (secular Hamiltonian). We then obtain the homonuclear Hamiltonians
\begin{equation}
  H_F=\frac{1}{2}\sum_{j<k}J^F_{jk}(2\sigma_z^j \sigma_z^{k}-\sigma_x^j \sigma_x^{k}-\sigma_y^j \sigma_y^{k}) \qquad H_P=\frac{1}{2}\sum_{\lambda<\kappa}J^P_{\kappa\lambda}(2s_z^\lambda s_z^{\kappa}-s_x^\lambda s_x^{\kappa}-s_y^\lambda s_y^{\kappa})
\end{equation}
and the heteronuclear interaction between the $F$ and $P$ spins,
\begin{equation}
  H_{FP}=\sum_{k,\kappa} J^{FP}_{k,\kappa}\sigma_z^ks_z^\kappa,
\end{equation}
with $J_{jk}=\hbar\gamma_j\gamma_k\frac{1-3\cos(\theta_{jk})^2}{4|\vec r_{jk}|^3}$, where $\theta_{jk}$ is the angle between the vector $\vec r_{jk}$ and the magnetic field $z$-axis. The maximum values of the couplings (for the closest spins) are given respectively by $J^F=-8.19$ krad s$^{-1}$, $J^P=0.30$ krad s$^{-1}$ and $J^{FP}=1.53$ krad s$^{-1}$. 
\begin{figure}[b]
\centering 
\includegraphics [width=0.8\linewidth ]{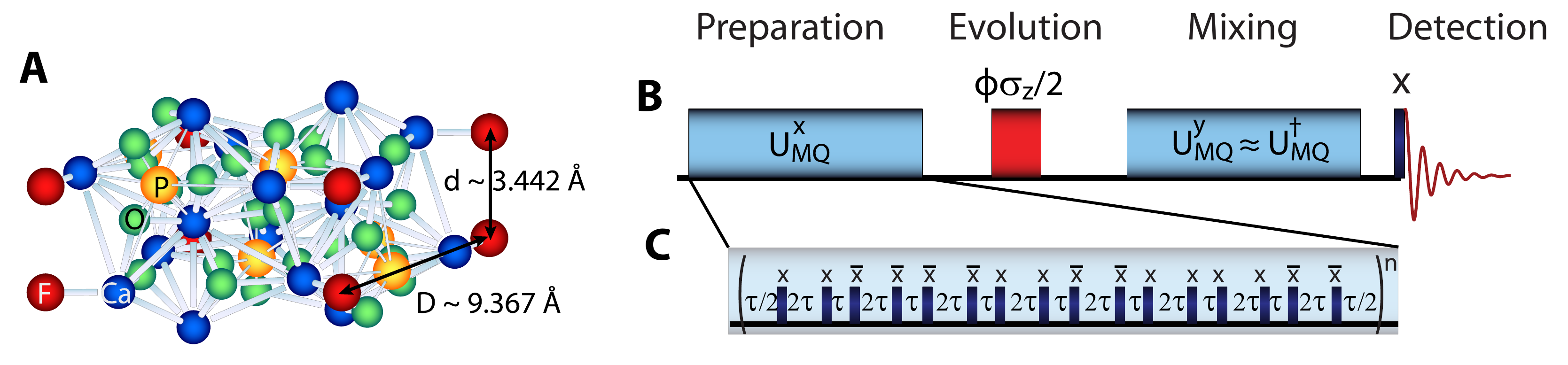}
\caption{\textbf{A} Fluorapatite crystal structure, showing the Fluorine and Phosphorus spins in the unit cell. \textbf{B} NMR scheme for the generation and detection of MQC. In the inset  (\textbf{C}) an exemplary pulse sequence for the generation of the double-quantum Hamiltonian. Note that thanks to the ability of inverting the sign of the Hamiltonian, the scheme amounts to measuring out-of-time order correlations.
}\label{fig:mqcd}
\end {figure}

The dynamics of this complex many-body system can be  mapped to a much simpler, quasi-1D system. First, we note that when the crystal is oriented with its $c$-axis parallel to the external magnetic field
the coupling of fluorine spins to the closest off-chain fluorine spin is $\approx40$ times weaker, while in-chain, next-nearest neighbor couplings are $8$ times weaker. 
 Previous studies on these crystals have indeed observed dynamics consistent with spin chain models, and the system has been proposed as solid-state realizations of quantum wires
~\cite{Cappellaro07l,Cappellaro11,Ramanathan11}. This approximation of the experimental system to a 1D, short-range system, although not perfect has been shown to reliably describe experiments for relevant time-scales~\cite{Rufeil-Fiori09b,Zhang09}. The approximation breaks down at longer time, with a convergence of various effects: long-range in-chain and cross chain couplings, as well as pulse errors in the sequences used for Hamiltonian engineering. In addition, the system also undergoes spin relaxation, although on a much longer time-scale ($T_1=0.8 s$ for our sample). 

In our experimental analysis, we mapped the physical system to the simple, nearest-neighbor 1D spin chain model. This allowed us to analyze the data with a well-known model, that furthermore leads to analytical solutions at least for the non-interacting case. However, it would be interesting to further analyze if the discrepancies from the ideal model, and in particular long-range interactions, give rise to further, interesting properties when exploring the systems for times where the approximation is no-longer as good. For example, the expected low growth of entanglement in long-range coupling systems is expected to exhibit a power-law~\cite{Pino14}, instead of a logarithmic growth, while the question of whether localization can be seen in long-range, higher-dimensional systems is still open.

\section{Entanglement/R\'enyi Entropy, Mutual Information, and average correlation metrics}
The many-body localized phase can be identified by the behavior of the entanglement entropy calculated on a subsystem. Unfortunately, this quantity is typically challenging to measure, and other metrics have been proposed to analyze the dynamics of out-of-equilibrium systems. 
Previous works have used the Hamming distance to quantify localization for pure states~\cite{Smith16, Hauke15}, but this metric shows quantitatively similar dynamics in both AL and MBL. Out-of-time ordered correlations (OTOC) between spatially separated local observables have been proposed as a metric to distinguish AL and MBL~\cite{Huang16}, and recently measured in some systems~\cite{Garttner16x,Li16x}. However, OTOC of local operators and other correlation metrics such the QFI are difficult to measure in large many-body systems, as they require the ability to address a subset of the total system. While the QFI has been linked to  the experimentally accessible dynamic susceptibility~\cite{Hauke16}, this relationship is valid only for systems at thermal equilibrium.
The metric we introduced in the main text directly aims at evaluating the spread of correlations, its saturation due to Anderson localization, and its slow, logarithmic growth in the presence of interaction (MBL). Here we compare our metric with other proposed metrics, to evaluate its robustness.
\subsection{Average Correlation Length and Distance}
In the main text we focused on the average correlation length as the chosen metric to experimentally detect the MBL phase. 
Here we introduce a second metric, that we call the \textbf{\textit{average correlation distance}} that can serve the same purpose and it is even more closely related to the notion of localization length. We then proceed to compare these two metrics to known measures of localization.

In the main text we decomposed the time-dependent density matrix using   operators \(\mathcal{B}^s_k\) composed of tensor products of $k$ Pauli matrices and $L-k$ identity operators. 
An alternative decomposition  is
\begin{align}
\rho(t)=\frac\openone{2^L}-\frac{\epsilon\sqrt{L}}{2^L} \sum_{k=1}^L\sum_{j=1}^{L+1-k} \sum_{r=1}^{\xi_k}a^{r}_{j,j+k-1}(t)\mathcal{A}^{r}_{j,j+k-1},
\label{eq:rho}
\end{align}
where $\mathcal{A}^{r}_{j,j+k-1}$ represents an operator composed of tensor products of Pauli matrices where the two farthest nonidentity operators are located at sites $j$ and $j+k-1$;
for each \(k\) there are \(\xi_k\) such configurations labeled by $r$. 

We can then define the  \textbf{\textit{average correlation distance}} over which spin correlations have established in the system:
\begin{align}
D_c=\sum_{k=1}^{L} k d_k
\label{eq:Dc}
\end{align}
where \(d_k=\sum_{j=1}^{L+1-k} \sum_{r=1}^{\xi_k} [a^r_{j,j+k-1}]^2\) 
is the contribution of all possible spin correlations 
over distance $k$, and satisfies the normalization \(\sum_{k=1}^L d_k =1\). \(D_c\) is a  measure of how far information has spread within the system.

We expect $D_c$ to have a qualitatively similar behavior to $L_c$, as indeed it is even more closely related to the notion of localization length.  However, measuring \(d_k\) and $D_c$ is  challenging, since for a generic many-body spin Hamiltonian the number of configurations \(\xi_k\) is exponential in \(k\) and \(L\). In addition, we cannot rely on the measurable MQC intensities to extract $d_k$. 
Indeed, since collective rotations cannot distinguish correlations such as \(\sigma_x^1 \sigma_x^2\) and \(\sigma_x^1 \sigma_x^5\), it is impossible to separate their contributions into $d_2$ and $d_5$, even if they can be correctly classified when measuring $f_2$. 
We find however that for  noninteracting systems, and for simple initial states such as \(\rho_\text{eq}\), all many-spin correlations are in the form \(\mathcal A_k=\mathcal B_k\sim\sigma_a^j (\prod_{l=j+1}^{k+j-2}\sigma_z^l) \sigma_b^{k+j-1}\), where the end spins \(\sigma_{a,b}\) are either \(\sigma_x\) or \(\sigma_y\). 
Then, in these systems a spin correlation established over distance $k$ corresponds to a correlation amongst $k$ spins, thus the average distance $D_c$  can be alternatively understood as the average number of correlated spins, that is, the average correlation length $L_c$.
While for noninteracting systems, $f_k=d_k$ and consequently $L_c=D_c$,  for interacting systems the two metrics are different, but they are  equally good correlation metrics in distinguishing MBL from AL.

\subsection{Comparison with known metrics of localization}
\begin {figure}
\centering 
\includegraphics[width=0.8\textwidth]{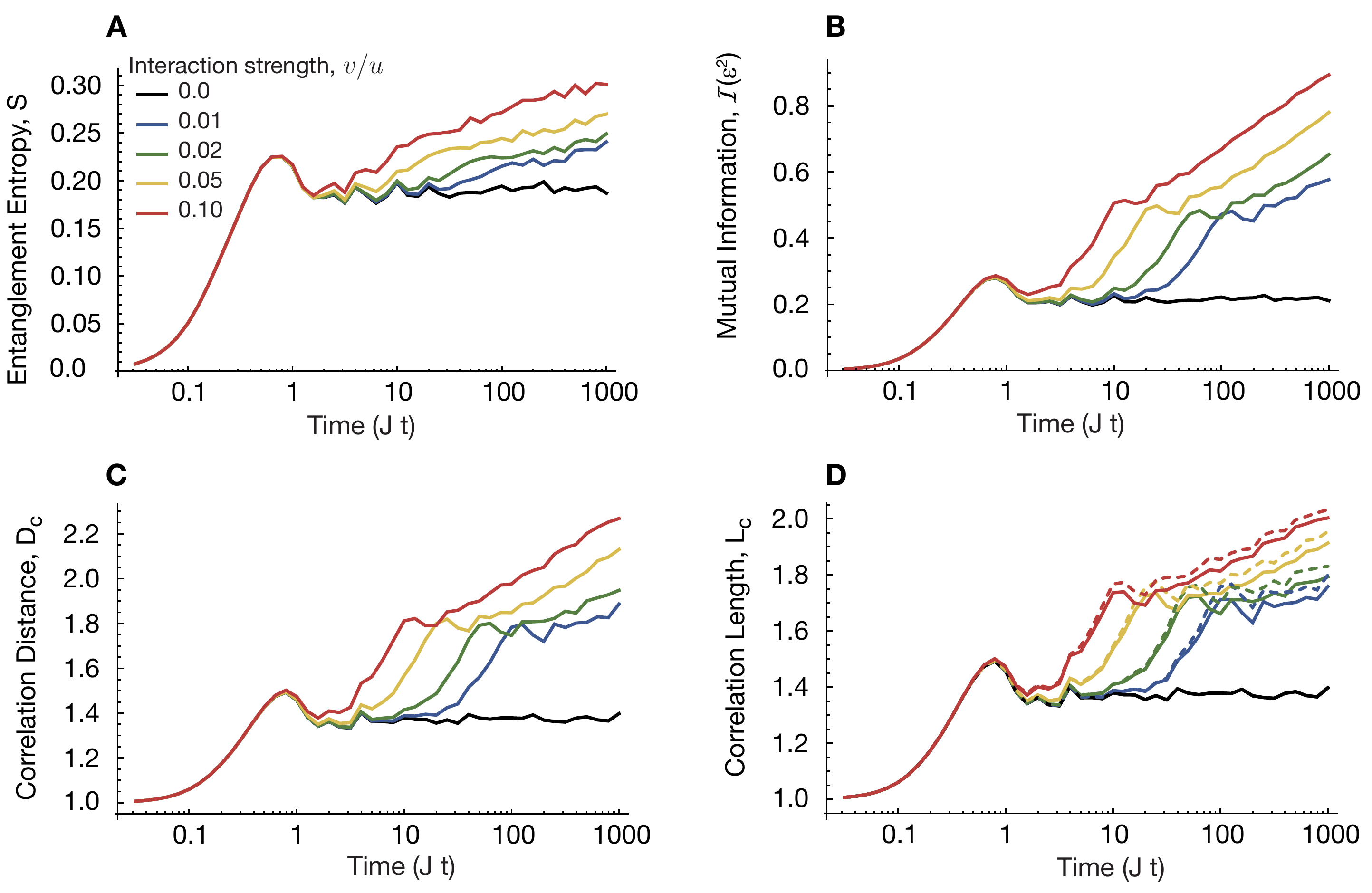}
\caption {Different correlation metrics used to distinguish MBL from AL. For pure states, the bipartite entanglement entropy is used (\textbf{A}). For $\rho_\text{eq}$, the mutual information (\textbf{B}), correlation distance (\textbf{C}), and correlation length (\textbf{D}) can be used. In \textbf{D}, the dashed lines corresponding to the approximated $L_c$ extracted from MQC intensities. All plots are for \(L=8\) with open boundary conditions, the disorder \(h_j\) is drawn uniformly from \([-W, W]\) with \(W=8\).
}
\label{fig:sim}
\end {figure}

In our experiments we engineered the Hamiltonian 
 \(H=uJ \sum_j(\sigma_+^j \sigma_+^{j+1}+\sigma_-^j \sigma_-^{j+1})+ vJ \sum_j(\sigma_+^j \sigma_-^{j+1}+\sigma_-^j \sigma_+^{j+1} - \sigma_z^j  \sigma_z^{j+1})+\sum_j h_j \sigma_z^j\) and showed how it leads to  an MBL state. Indeed, our collective control prevented us from engineering  Hamiltonians more commonly found in the literature. 
As this Hamiltonian has not been studied previously, here we show that its interacting version does indeed 
support MBL by calculating the bipartite entanglement entropy for an \(L=8\) chain averaged over \(10^3\) disorder realizations (Fig.~\ref{fig:sim}.\textbf{A}). We use the initial pure product state \(|\uparrow \uparrow \uparrow \cdots \uparrow \uparrow\rangle\), and calculate the entropy  \(S=-\text{Tr}[\rho_L\log\rho_L]\), where \(\rho_L\) is the reduced density matrix of the left half of the chain. We observe a logarithmic growth of entanglement in the presence of weak interactions, a defining feature that separates MBL from AL.

Next, consider our high-temperature system. As the largest component of the Hamiltonian is the Zeeman interaction, $H_Z=\omega_0Z/2$, with $\omega_0=(2\pi)283$ MHz, the thermal equilibrium state can be well approximated by $\rho_\text{eq}=e^{-\beta H}\approx e^{-\beta \omega_0Z/2}$. In the experimental conditions, $\epsilon= \beta\omega_0/2=\gamma_FB/2k_BT\approx 2.3\times10^{-5}$ and we have so far considered an expansion to first order in $\epsilon$, $\rho_\text{eq}=\frac1{2^L}\left(\openone-\epsilon Z\right)$. 
Since the entropy of the reduced density matrix  is zero to first order in $\epsilon$, we consider the expansion up to second order,
\[\rho_\text{eq}=2^{-L}[(1-\epsilon^2 L/2)\openone-\epsilon Z +\epsilon^2 Z^2/2]\]
To second order in $\epsilon$, the entropy of the reduced density matrix of the right half of the chain is 
\[S_R=-\text{Tr}[\rho_R\log(\rho_R)]\approx \frac12 \left(L\log(2)-\frac{\epsilon^2}{2^{3L/2}}\text{Tr}[\delta\rho_R^2(t)]\right),\] 
where $\delta\rho_R(t)=\text{Tr}_L[U(t)ZU(t)^\dag]$. 
In order to avoid the large constant term in this expression, we can calculate instead  the mutual information (MI) $\mathcal{I}$, defined as $\mathcal{I}=S_L+S_R-S_{L\cup R}$~\cite{Groisman05}.  
MI is a measure of the total correlations, both quantum and classical, in the system. Notice that MI reduces to twice the bipartite entanglement entropy for pure states, since $S_{L\cup R}=0$ and $S_L=S_R$ for pure states. For $\rho_\text{eq}$, MI can be expressed as
\begin{align}
\mathcal{I}=\frac{\epsilon^2 }{2}\left(L-{2^{-3L/2}}\left[\text{Tr}(\delta\rho_L^2)+\text{Tr}(\delta\rho_R^2)\right]\right)
\label{eq:mi}
\end{align} 
Interestingly, $\mathcal{I}$ also saturates for noninteracting systems and increases logarithmically when interactions is added (see Fig.~\ref{fig:sim}.\textbf{B}), with a similar behavior as $S_R$. In the main text, we plot 
$1-2^{-\frac L2}\frac{2}{L}\text{Tr}[\delta\rho_R^2(t)]=\frac{2^{L+1}}{\epsilon^2L/2}[S_R-\frac{L}2 \log (2)]+1$ in Fig.~2.

Finally we show the dynamics of $D_c$ in Fig.~\ref{fig:sim}.\textbf{C}, and compare $L_c$ with the approximated $L_c$ (using the method explained in section~\ref{sec:fermions} and \ref{sec:MQC}) in Fig.~\ref{fig:sim}.\textbf{D}. These correlation metrics all display logarithmic growth in the MBL phase, and thus can be applied analogously as entanglement entropy in distinguishing MBL from AL for highly mixed interacting systems.

\subsection{Relation with  Mutual Information and Entanglement/R\'enyi entropy}
Here we argue that $D_c,\ L_c$ and $\mathcal{I}$ are closely related when considering highly mixed states such as $\rho_\text{eq}$. We show this relationship using $D_c$, but a similar argument also holds for $L_c$. 
Since the total weight of all correlations sums to one, we can rewrite the last term in Eq.~(\ref{eq:mi}) as
\begin{align}
\text{Tr}(\delta\rho_L^2)+\text{Tr}(\delta\rho_R^2)=2^{3L/2}L\left(1-\sum_{j=1}^{L/2} \sum_{k=L/2+1}^{L} \sum_{r=1}^{\xi_{k+1-j}} \left[a^r_{j,k}\right]^2\right)
\label{eq:mi_1}
\end{align}
The LHS of Eq.~(\ref{eq:mi_1}) represents the total weight of correlations contained in the left and right halves of the chain. It is different from one by the  amount corresponding to the correlations across the middle of the chain. This difference is the last term inside the bracket on the RHS. If we approximate the weights of local correlations as averages of bulk weights, i.e., $\sum_{r=1}^{\xi_2} [a^r_{L/2,L/2+1}]^2=\frac{d_2}{L-1}$, $\sum_{r=1}^{\xi_3} [a^r_{L/2,L/2+2}]^2=\frac{d_3}{L-2}$, and etc., we can place a lower bound on this difference as
\begin{align}
\sum_{j=1}^{L/2} \sum_{k=L/2+1}^{L} \sum_{r=1}^{\xi_{k+1-j}} \left[a^r_{j,k}\right]^2=\sum_{k=2}^{L/2+1}\frac{k-1}{L-(k-1)}d_k+\sum_{k=L/2+2}^L d_k \geq \sum_{k=1}^L\frac{k-1}{L-1}d_k =\frac{D_c -1}{L-1}
\label{eq:ineq}
\end{align} 
The approximation $\sum_{r=1}^{\xi_k} [a^r_{L/2,L/2+k}]^2\approx\frac{d_k}{L-k}$ becomes exact for a translationally invariant system, which is a good approximation for a large system with weak disorder. Similarly, the inequality in the last step of Eq.~(\ref{eq:ineq}) becomes tighter for larger systems. Even in the presence of strong disorder, the approximation is still valid upon averaging over many disorder realizations.
Substituting Eq.~(\ref{eq:ineq}) into Eq.~(\ref{eq:mi}) we obtain a lower bound on the mutual information in terms of the average correlation distance:
\begin{align}
\mathcal{I} \geq \frac{\epsilon^2 L}{2(L-1)}(D_c-1)  
\end{align}
With a similar reasoning, we can obtain $\mathcal{I} \geq \frac{\epsilon^2 L}{2(L-1)}(L_c-1)$.  
In order to relate $L_c$ and $D_c$ to entanglement entropy, we first note that $S_L\approx S_R$ after disorder averaging. This allows us to write
\begin{align*}
S_L=\frac{L}{2}\left(\log(2)-\frac{\epsilon^2}{2}\right)+\frac{\mathcal{I}}{2}
\end{align*}
Using Eq.~(\ref{eq:ineq}), we can place a lower bound on the time-dependent part of the entanglement entropy as 
\begin{align}
\frac{\epsilon^2L}{4(L-1)}(D_c-1)\leq S_L-\frac{L}{2}\left(\log(2)-\frac{\epsilon^2}{2}\right)
\end{align}
Similarly for the second order R\'enyi entropy: $S_L^{(2)}=\frac{L}{2}\log(2)-\frac{\epsilon^2}{2^{3L/2}}\text{Tr}(\delta\rho^2_L)$, the time-dependent component has a lower bound given by
\begin{align*}
\frac{\epsilon^2L}{2(L-1)}(D_c-1)\leq S_L^{(2)}-\frac{L}{2}(\log(2)-\epsilon^2)
\end{align*}

Beyond the bounds described above, we can obtain explicit expressions for  the second order R\'enyi entropy (or other entanglement metrics) that can be related to measurable quantities in our system. Consider for example the state expansion in terms of the $\mathcal B_k^r$ operators, as in Eq.~(\ref{eq:rhoN}). Taking the partial trace of these traceless operators, we obtain the R\'enyi entropy
for the left half of the spin chain:
\begin{equation*}
S_L^{(2)}=\frac L2\log(2)-\frac{\epsilon^2}{2^{3L/2}}\sum_{k=2}^{L/2}\sum_{s}^{\zeta^L_k} (b_{k}^s)^2,	
\end{equation*}
where $\zeta^L_k$ counts only the configurations in the left half of the chain. We note that most generally, for $k>1$,  we have $\zeta^L_k=\frac{L/2-k+1}{L-k+1}\zeta_k$, where $\zeta_k$ are all the $k$-spin configurations in the whole chain.  

Given the symmetry of the system, we expect that for a translationally invariant system, $b_{k}^r$ do not depend on the first spin position, and thus we have $\sum_{s}^{\zeta^L_k} (b_{k}^s)^2=(\zeta_k^L/\zeta_k) f_k$. While this is not exactly true for finite systems with open boundaries (and generally in the presence of disorder),  we can still extract an approximated R\'enyi entropy:
\begin{equation}
  S_L^{(2)}\approx\!\frac L2\log(2)-\frac{\epsilon^2}{2^{3L/2}}\left(\frac{f_{1}}2+2\sum_{k=2}^{L/2}\frac{L/2-k+1}{L-k+1}f_{k}\right)
\end{equation}
This approximated R\'enyi entropy also shows a distinctive logarithmic growth for the MBL state and can be measured experimentally with our MQC intensity protocol. It is thus an equivalent metric to $L_c$,  however we prefer to still refer to $L_c$ since it has a simpler physical interpretation.

\section{Measuring MQC Intensities}
\label{sec:MQC}
The complex dynamics of many-spin correlations can be partially elucidated by measuring MQC intensities. 
Quantum coherence of order \(k\) describe the contribution of terms \(|m_{\bf n}\rangle\!\langle m'_{\bf n}|\) in the density matrix such that \(m_{\bf n}-m'_{\bf n}=k\), with \(m_{\bf n}\) the collective \(\sigma_{\bf n}\) eigenvalue (\({\bf n}\) here denote direction).  Quantum coherences in the \({\bf n}\)-basis are then classified based on their response to rotations around the \({\bf n}\) axis: A state of coherence order \(q\), when rotated around the axis \({\bf n}\) by an angle \(\varphi\), will pick up a phase equal to \(q\varphi\), i.e, \(e^{-i\varphi/2\sum_j\sigma_{\bf n}^j}\rho_q e^{i\varphi/2\sum_j\sigma_{\bf n}^j}=e^{-iq\varphi}\rho_q\). This property is often used in NMR experiments to select a particular coherence order, by a procedure  called \textit{phase cycling}~\cite{Bodenhausen84} that amounts to averaging measurements done with phase-shifted pulse sequences.

To see this, consider expanding the density matrix as a sum of quantum coherences: \(\rho=\sum_q \rho_q\). If we wish, for example, to keep only the even order coherences, we can measure \(\rho+\rho_\pi\), where \(\rho_\pi\) is the density matrix rotated around \(\bf n\) by \(\pi\). 
In \(\rho_\pi\), all odd order coherences pick up a minus sign and exactly cancel their counterparts when added to \(\rho\), leaving only \(\rho_{q\in\text{even}}\). We emphasize that quantum coherences are defined with respect to a given axis of rotation. Different axes of rotation give rise to different sets of quantum coherences.

Associated with the decomposition of a density matrix into quantum coherence is the concept of multiple quantum coherence (MQC) intensities. A MQC intensity of order \(q\) is defined as \(I_q=\text{Tr}[\rho_q \rho_{-q}]\), that is, \(I_q\) can be understood as the weight of the \(q\)-th order coherence in the density matrix. MQC intensities are an incomplete measure of many-spin correlations since a signal in \(I_q\) indicates there are at least \(|q|\) spins present in the correlations. On the other hand, a correlation with \(m\) spins can in principle give rise to all \(I_q\) with \(q=-m,-m+1,\cdots, m\). 

 MQC intensities can be measured in four steps. The schematic of a conventional MQC experiment is shown in Fig.~\ref{fig:mqcd}. 
During preparation, the system initially at \(\rho_\text{eq}\) is driven to evolve under a many-body Hamiltonian \(H\), thereby generating many-spin correlations. 
An encoding pulse, described by the unitary operator \(e^{-i\varphi Z/2}\), tags the quantum coherences according to their coherence orders. 
The refocusing step implements the time-reversed evolution to bring the system back into magnetization, which can then be detected by a \(\pi/2\) pulse. 
In our system, the sign of the Hamiltonian can be inverted by adjusting parameters in the pulse sequence (see  section~\ref{sec:AHT}). 
The overall signal of the MQC experiment can be expressed as
\begin{align}
S_\varphi&=\text{Tr}[U^\dagger e^{-i\varphi Z/2}U \rho_\text{eq} U^\dagger e^{i\varphi Z/2}U Z] \nonumber \\
&\propto\text{Tr}[e^{-i\varphi Z/2}\delta\rho e^{i\varphi Z/2} \delta\rho]=\sum_q e^{-i q\varphi}\text{Tr}[\delta\rho_q \delta\rho_{-q}]=\sum_q e^{-i q \varphi} I_q
\end{align}
where \(\delta\rho=UZU^\dagger \propto U\rho_\text{eq}U^\dagger -\openone/2^L\) and we have used the identity \(\text{Tr}[\delta\rho_q \delta\rho_p]=\delta_{q,-p}\) in the second to last step. 
To extract the intensities \(I_q\), we perform a series of MQC experiments as we vary \(\varphi\) from 0 to 2\(\pi\) in steps of \(\frac{\pi}{M}\), where \(M\) is the maximum coherence order to be measured. 
By performing a discrete Fourier transform (DFT) with respect to \(\varphi\), the MQC intensities can be found 
\begin{align}
I_q\propto\sum_{m=1}^{2M} e^{-i\frac{q m \pi}{M}} S_m,
\end{align}
where \(S_m\) is the signal of the \(m\)-th MQC experiment with \(\varphi =m\pi /M\). 
As explained in the main text, MQC intensities encoded in the \(\bf z\) axis cannot reveal the extent of spin-correlations generated in \(\rho_\text{eq}\) in our experiments. In the following sections we will explain how conventional MQC experiments can be modified to reveal the full extent of spin-correlations in spin chains.

Here we note that the schemes for measuring MQC intensities amounts to the detection of out-of-time order correlation (OTOC)~\cite{Garttner16x,Fan16x}.
Indeed, consider the operators $V\equiv Z=\sum_j\sigma_z^j$ and $W\equiv\Phi=e^{-i\phi Z/2}$. The signal $S_\phi$ corresponds to measuring the OTOC, $S_\phi(t)=\langle W(t)^\dag V^\dag(0)W(t) V(0)\rangle$, for a system at infinite temperature~\cite{Fan16x,Li16x}. While we cannot measure a full basis for a subsystem of the Hilbert space (which has been shown to yield the  second R\'enyi entropy~\cite{Fan16x}), as we have shown in the main text we can use the measured OTOC's to extract an equivalent metric of the the many-body localized phase. We expect that our metric could be of interest in studying many-body phase transitions or chaotic systems.

\section{Fermionic solution to noninteracting systems}
\label{sec:fermions}
In order to show how \(f_k\) and \(L_c\) can be extracted from MQC intensities, we need to first present a microscopic description of the spin-correlations generated by the Hamiltonian. 
In the main text, we introduce the generic (Floquet) Hamiltonian that we can experimentally generate with our control (see Eq.~(1)),
\begin{equation}
H=u  \sum_{j=1}^{L-1}\frac J2(\sigma_x^j \sigma_x^{j+1}-\sigma_y^j \sigma_y^{j+1})+b\sum_{j=1}^{L} \sigma_z^j  
+g \sum_{j=1}^{L} h_j \sigma_z^j+ v  \sum_{j=1}^{L-1}\frac J2(\sigma_x^j \sigma_x^{j+1}+\sigma_y^j \sigma_y^{j+1} - 2\sigma_z^j  \sigma_z^{j+1}).
\end{equation}
We can rewrite the spin Hamiltonian \(H\) in terms of fermion operators using a Jordan-Wigner transformation~\cite{Jordan28}: \(c_j= \sigma_-^k\prod_{k<j}\sigma_z^j\),
\begin{align}
H=&-u J  \sum_{j=1}^{L-1}(c_j^\dag c_{j+1}^\dag+c_{j+1}c_j)+b\sum_{j=1}^{L} (2c_j^\dagger c_j-1) +g \sum_{j=1}^{L} h_j (2c_j^\dagger c_j- 1) \nonumber \\
& - v J  \sum_{j=1}^{L-1}(c_j^\dag c_{j+1}+c_{j+1}^\dag c_j) -  v J  \sum_{j=1}^{L-1} (2c_j^\dagger c_j- 1)(2c_{j+1}^\dagger c_{j+1}- 1).
\label{eq:fermeq}
\end{align}
In particular, this makes it apparent that only the last term of the equation corresponds to interactions.

\begin {figure}
\centering 
\includegraphics[width=0.8\textwidth]{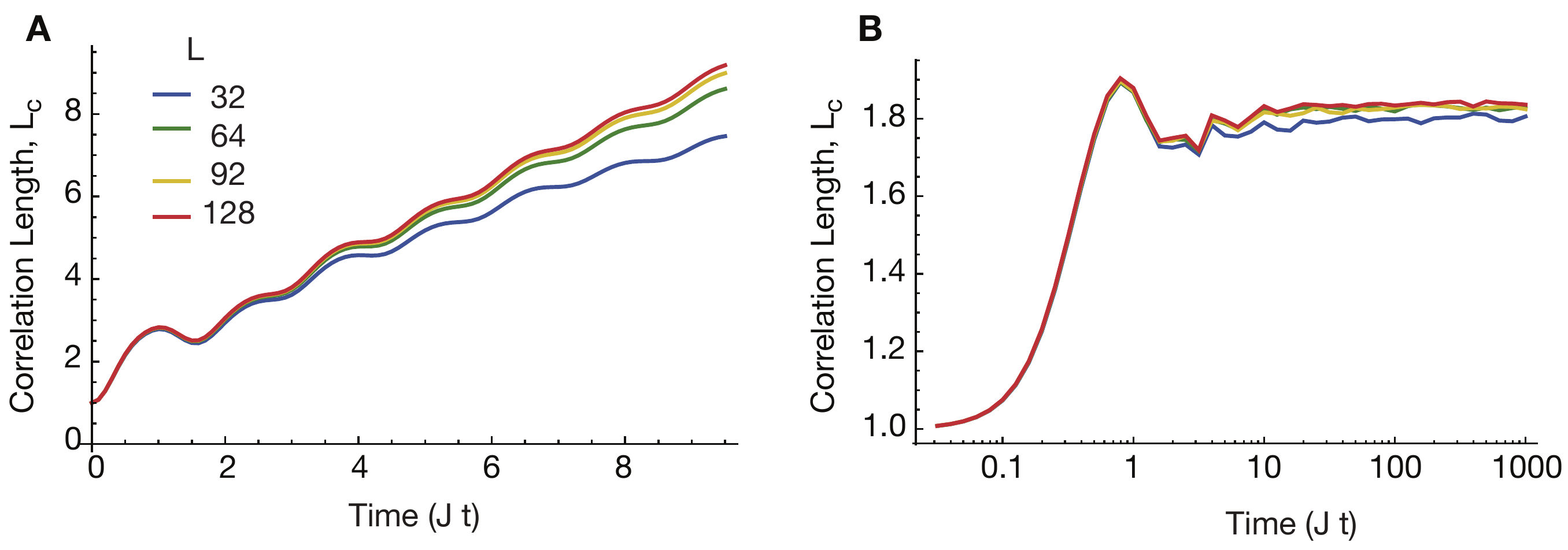}
\caption {The dynamics of $L_c$ under uniform (\textbf{A}) or disordered field (\textbf{B}) for different system sizes. With disordered fields, AL causes $L_c$ to saturate at a value independent of $L$. With uniform field, $L_c$ increases faster for larger $L$, indicating that the system is delocalized in the thermodynamic limit.
}
\label{fig:lc_length}
\end {figure}

The dynamics of the noninteracting Hamiltonian (ignoring $\sum_j\sigma_z^j \sigma_z^{j+1}$ term) \(H=u J \sum_j(\sigma_+^j \sigma_+^{j+1}+\sigma_-^j \sigma_-^{j+1})+ v J \sum_j(\sigma_+^j \sigma_-^{j+1}+\sigma_-^j \sigma_+^{j+1})+g\sum_j h_j \sigma_z^j + b\sum_j \sigma_z^j\) can be solved by mapping the system into a chain of spinless fermions~\cite{Jordan28,Feldman96,Cappellaro07l}. The time-dependent component of the initial density matrix $\rho_\text{eq}$ evolves as 
\begin{align}
\delta\rho/\sqrt{L}=\sum_j \mu_{jj}\sigma_z^j+\sum_{j,k>j}\left[-(\mu+\eta)_{jk}\sigma_x^j \sigma_x^k-(\mu-\eta)_{jk} \sigma_y^j \sigma_y^k +(\nu+\chi)_{jk}\sigma_y^j \sigma_x^k+(\nu-\chi)_{jk}\sigma_x^j\sigma_y^k\right]\prod_{j<l<k} \sigma_z^l,
\label{eq:rhofull}
\end{align}
where \(\mu\) is real and symmetric, whereas \(\chi\), \(\eta\), \(\nu\) are real and antisymmetric; they correspond to the four possible ways to correlate spins in the noninteracting system. 
In terms of these matrices \(f_k\) can be expressed as
\begin{align}
f_1=\bar{\mu}_0, \qquad f_{k>1}=2(\bar{\mu}_{k-1}+\bar{\chi}_{k-1}+\bar{\eta}_{k-1}+\bar{\nu}_{k-1}),
\label{eq:eqfk}
\end{align}
where \(\bar{\mu}_k=\sum_j \mu_{jj+k}^2\), and similar expressions hold for \(\chi\), \(\eta\), and \(\nu\). 
Once \(f_k\) is found, the correlation length can be calculated as \(L_c=\sum_k k f_k\). Notice that in the noninteracting system the average correlation length is exactly equal to the average correlation distance, i.e., $f_k=d_k$ and \(L_c=D_c\). Analytical expressions for these two quantities can be obtained in some limiting cases. 
For instance, let $u=1$, $v=g=0$, and consider the double quantum Hamiltonian: \(H_\text{dq}=J\sum_j (\sigma_+^j \sigma_+^{j+1}+\sigma_-^j \sigma_-^{j+1})+b\sum_j \sigma_z^j\). 
We can rewrite \(H_\text{dq}\) in terms of fermion operators using the Jordan-Wigner transformation,
\begin{align*}
H_\text{f}=-J \sum_j(c_j^\dagger c_{j+1}^\dagger +c_{j+1} c_j)+b\sum_j (c_j^\dagger c_j- c_j c_j^\dagger),
\end{align*}
where the fermion operators satisfy \(\{c_j^\dagger, c_k\}=\delta_{jk}\), \(\{c_j, c_k\}=\{c_j^\dagger, c_k^\dagger\}=0\). Next we perform a modified Fourier transformation given by \(c_j^\dagger=(-i)^j\sqrt{\frac{2}{L+1}}\sum_q \sin \left(\frac{j q\pi}{L+1}\right)d_q^\dagger\), and write the Hamiltonian in momentum space as
\begin{gather*}
H_\text{f}=\sum_q 
\begin{pmatrix}
d^\dagger_q & d_{\bar{q}}
\end{pmatrix}
\begin{pmatrix}
b & -i J_q \\
i J_q & -b
\end{pmatrix}
\begin{pmatrix}
d_q \\
d^\dagger_{\bar{q}} \\
\end{pmatrix},
\end{gather*}
where \(\bar{q}=L+1-q\), and \(J_q=J\cos\left( \frac{q\pi}{L+1}\right)\). Instead of using a Bogoliubov transformation to diagonalize the matrix, we use Heisenberg's equation of motion to directly obtain the dynamics of each pair of modes,
\begin{gather*}
\begin{pmatrix}
d_q(t) \\
d^\dagger_{\bar{q}}(t) \\
\end{pmatrix}
=\exp[-2i(b \sigma_z +J_q \sigma_y)t]
\begin{pmatrix}
d_q \\
d^\dagger_{\bar{q}} \\
\end{pmatrix}
=
\begin{pmatrix}
\cos (2\omega_q t)-i \cos\theta_q \sin(2\omega_q t) & -\sin\theta_q \sin (2\omega_q t) \\
\sin\theta_q \sin (2\omega_q t) & \cos(2\omega_q t)+i \cos\theta_q \sin (2\omega_q t)
\end{pmatrix}
\begin{pmatrix}
d_q \\
d^\dagger_{\bar{q}} \\
\end{pmatrix}
\end{gather*}
where \(\omega_q=\sqrt{b^2+J_q^2}\), \(\cos\theta_q=b/\omega_q\), and \(\sin\theta_q=J_q/\omega_q\). Note that $d_q$ represents a fermion operator in momentum space, it is not to be confused with $d_k$ in Eq.~(\ref{eq:Dc}). In the thermodynamic limit (\(L\rightarrow \infty\)), the correlation length in the absence of any transverse field (\(b=0\)) can be expressed as
\begin{align}
L_c=\sum_{k=1}^\infty k f_k=1+16J^2 t^2[\mathcal{J}_0^2(4Jt)+\mathcal{J}_1^2(4Jt)]-4Jt\mathcal{J}_0(4Jt)\mathcal{J}_1(4Jt),
\end{align}
where \(\mathcal{J}_k\) is the \(k\)-th Bessel function of the first kind. At long times \(Jt \gg 1\), \(L_c\) has the asymptotic form \(L_c\sim 8Jt/\pi\). This shows that the average spread of correlations in the noninteracting spin chain has a light-cone like behavior. When \(b\neq 0\), \(L_c\) cannot be expressed analytically. In the regime \(b\ll J\), we can approximate \(L_c\) as a series expansion in powers of \(J/b\). The first two terms are given by
\begin{align}
L_c=1+\frac{J^2}{b^2}\left[\frac{1}{2}+\frac{1}{2}\mathcal{J}^2_0\left(\frac{J^2t}{b}\right)-\mathcal{J}_0\left(\frac{J^2t}{b}\right)\cos\left(4bt+\frac{J^2t}{b}\right)+\mathcal{J}_1\left(\frac{J^2t}{b}\right)\sin\left(4bt+\frac{J^2t}{b}\right)\right. \nonumber \\
\left. +\frac{J^4t^2}{b^2}\left(\mathcal{J}^2_0\left(\frac{J^2t}{b}\right)+\mathcal{J}^2_1\left(\frac{J^2t}{b}\right)\right)-\frac{J^2 t}{b}\mathcal{J}_0\left(\frac{J^2t}{b}\right)\mathcal{J}_1\left(\frac{J^2t}{b}\right) \right]+O\left(\frac{J^4}{b^4}\right),
\end{align}
which has the asymptotic behavior \(L_c\sim \frac{2 J^2 t}{\pi|b|}\left[\frac{J^2}{b^2}+O\left(\frac{J^4}{b^4}\right)\right]\) at long timescales. In Fig.~\ref{fig:lc_length} we show the dynamics of \(L_c\) in the presence of either disorder or uniform transverse field for different system sizes. It is clear that in the case of disordered fields, AL causes \(L_c\) to saturate at the localization length \(\xi\), which is independent of \(L\) if \(\xi\ll L\) (Fig.~\ref{fig:lc_length}.\textbf{B}). On the other hand, \(L_c\) does not saturate with transverse field, but instead grows with a reduced velocity (Fig.~\ref{fig:lc_length}.\textbf{A}). Notice that for short times, the oscillations induced by transverse field lead to an apparent localization in the system. 

\section{Extracting $L_c$ From MQC}\label{sec:MQC}
Thanks to the structure of the density matrix shown in Eq.~(\ref{eq:rhofull}), we can design a protocol for extracting the average correlation length for the non-interacting system, and to extract a close approximation in the case of interactions. 
The first step is to extract signal separately from each of the four sectors, \(\bar{\mu}_k\), \(\bar{\chi}_k\), \(\bar{\eta}_k\), and \(\bar{\nu}_k\), defined above in Eq.~(\ref{eq:eqfk}). We use time reversal of the evolution Floquet Hamiltonian to measure $\rho(t)$ and $\rho(-t)$. Since \(\rho(-t)=\rho^*(t)\), we can thus obtain the real and imaginary parts of \(\rho\) from  \(\rho(t)\pm\rho(-t)\).

The real part, \(\text{Re}[\rho]\), only contains correlations with an even number of \(\sigma_y\) and $\sigma_x$ operators, i.e, operators such as \(\mathcal B^{(\textrm{re})}= \sigma_x^i\prod_{k=i+1}^{j-1}\sigma_z^k \sigma_x^{j}\pm\sigma^i_y\prod_{k=i+1}^{j-1}\sigma_z^k \sigma^j_y\). The imaginary part, instead, \(\text{Im}[\rho]\), is composed of $\mathcal B^{(\textrm{im})}$ operators   with \(\sigma_x, \sigma_y\) operators (and vice-versa) as end spins. 
We can then further subdivide 
\(\text{Re}[\rho]\) and \(\text{Im}[\rho]\) using phase cycling~\cite{Bodenhausen84}.
We note that \(\sigma_x^i\dots\sigma_x^j+\sigma_y^i\dots\sigma_y^j\) and \(\sigma_x^i\dots\sigma_y^j-\sigma_y^i\dots\sigma_x^j\) are invariant under rotations around \(z\) whereas \(\sigma_x\dots\sigma_x-\sigma_y\dots\sigma_y\) and \(\sigma_x\dots\sigma_y+\sigma_y\dots\sigma_x\) pick up a minus sign when rotated by \(\pi/2\). 
Thus, by measuring $\text{Re}[\rho]\pm U(\pi/2)\text{Re}[\rho]U(\pi/2)^\dag$ (and similarly for the imaginary part) we can finally find the contributions from the four sectors. 

Next we measure the MQC intensities encoded in the \(x\) axis for each of these four sectors. 
For the non-interacting system, the MQC intensities encoded along the \(x\) axis for each sector (labeled by the superscript) can be written as
\begin{equation}
\begin{aligned}
I_q^{\mu}&=\frac{\delta_{1|q|}}{2}\bar{\mu}_0 + \sum_{k=1}\left[
\frac{1}{2^{k+1}}
\begin{pmatrix}
k+1 \\
\frac{k+1-q}{2}\\
\end{pmatrix}
+ \frac{1}{2^{k-1}}
\begin{pmatrix}
k-1 \\
\frac{k-1-q}{2}\\
\end{pmatrix}\right]
\bar{\mu}_k   \\
I_q^{\eta}&=\sum_{k=1}\left[
\frac{1}{2^{k+1}}
\begin{pmatrix}
k+1 \\
\frac{k+1-q}{2}\\
\end{pmatrix}
+ \frac{1}{2^{k-1}}
\begin{pmatrix}
k-1 \\
\frac{k-1-q}{2}\\
\end{pmatrix}\right] 
\bar{\eta}_k  
\end{aligned} \qquad
\begin{aligned}
I_q^{\chi}&=
\sum_{k=1}\frac{1}{2^{k-1}}
\begin{pmatrix}
k \\
\frac{k-q}{2}\\
\end{pmatrix}\bar{\chi}_k \\
I_q^{\nu}&=
\sum_{k=1}\frac{1}{2^{k-1}}
\begin{pmatrix}
k \\
\frac{k-q}{2}\\
\end{pmatrix}\bar{\nu}_k
\end{aligned}
\label{eq:eqmqc}
\end{equation}
\(I_q^\mu\) and \(I_q^\eta\) are defined for \(k-q\in \text{odd}\), whereas \(I_q^\chi\) and \(I_q^\nu\) are defined for \(k-q \in \text{even}\). These expressions amount to a linear transformation \(f_k=\sum_{jk} M^{(j)}_{kq}I^j_q\), where \(M^{(j)}\) are constant matrices calculated from the inverse of Eq.~(\ref{eq:eqmqc}).
Notice that the MQC spectrum is symmetric, \(I_q=I_{-q}\), and satisfies the normalization condition \(\sum_{q}I_q=1\). 
All experimental data presented in the main text have been normalized accordingly. By inverting Eq.~(\ref{eq:eqmqc}) we can extract \(\bar{\mu}_k\), \(\bar{\chi}_k\), \(\bar{\eta}_k\), and \(\bar{\nu}_k\) from measured MQC intensities and find \(f_k\) using Eq.~(\ref{eq:eqfk}). 

\begin{figure}[ht]
\centering 
\includegraphics[width=0.9\textwidth]{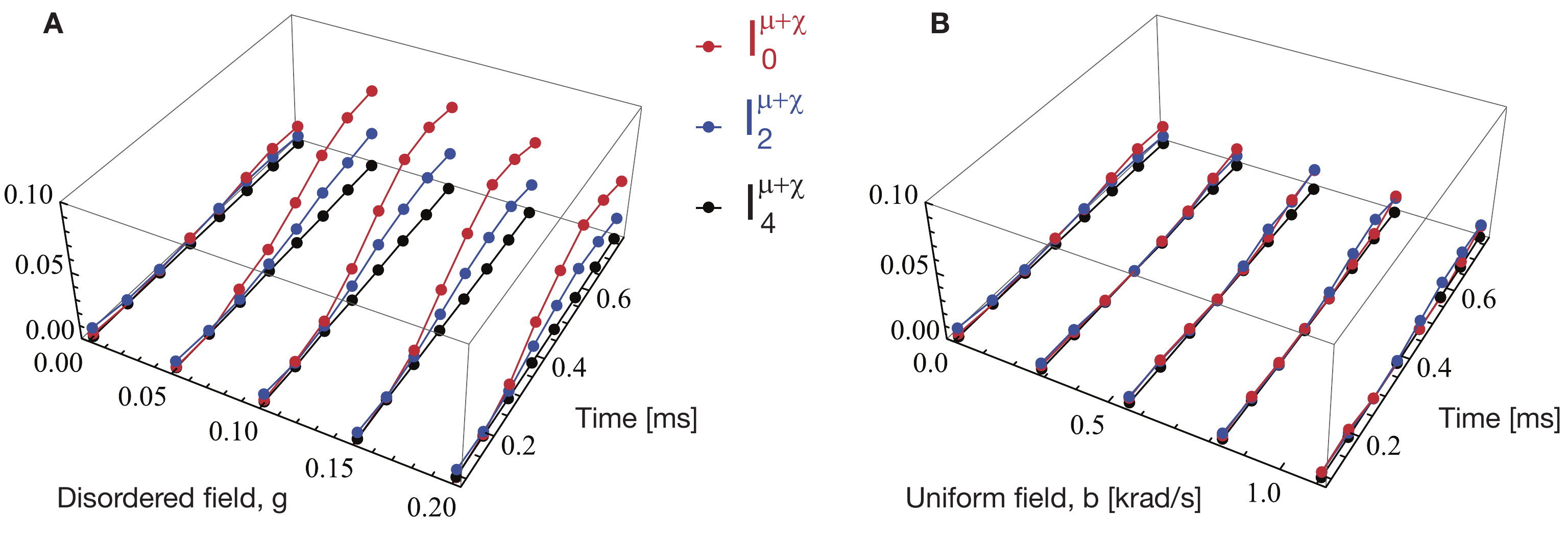}
\caption {Experimentally measured MQC intensities in the $\mu+\chi$ sector for different strength of disorder (\textbf{A}) and transverse uniform field (\textbf{B}). In the absence of any disorder, it can be shown $I^{\mu+\chi}=0$ based on symmetry arguments. The $g=0.15$ and $b=0.826$ krad/s data are compared and shown in the main text.
}
\label{fig:chi}
\end{figure}

In the case of the double quantum Hamiltonian, \(H_\text{dq}=J\sum_j (\sigma_+^j \sigma_+^{j+1}+\sigma_-^j \sigma_-^{j+1})+g\sum_j h_j \sigma_z^j+b\sum_j \sigma_z^j\), there is a symmetry given by \([\sum_j (-)^j \sigma_z^j, H_\text{dq}]=0\). For \(\rho_\text{eq}\) this symmetry leads to \(\bar{\mu}_{k\in\text{odd}}=\bar{\chi}_{k\in\text{odd}}=\bar{\eta}_{k\in\text{even}}=\bar{\nu}_{k\in\text{even}}=0\). This simplifies the MQC experiments considerably, and we can extract all the non-vanishing coefficients by decomposing the density matrix into two sectors instead of four
\begin{equation}
\begin{aligned}
I_{q\in \text{odd}}^{\mu+\chi}&=\frac{\delta_{1|q|}}{2}\bar{\mu}_0 + \sum_{k=2,4,\cdots}\left[
\frac{1}{2^{k+1}}
\begin{pmatrix}
k+1 \\
\frac{k+1-q}{2}\\
\end{pmatrix}
+ \frac{1}{2^{k-1}}
\begin{pmatrix}
k-1 \\
\frac{k-1-q}{2}\\
\end{pmatrix}\right]
\bar{\mu}_k  \\
I_{q\in \text{even}}^{\eta+\nu}&=\sum_{k=1,3,\cdots}\left[
\frac{1}{2^{k+1}}
\begin{pmatrix}
k+1 \\
\frac{k+1-q}{2}\\
\end{pmatrix}
+ \frac{1}{2^{k-1}}
\begin{pmatrix}
k-1 \\
\frac{k-1-q}{2}\\
\end{pmatrix}\right]
\bar{\eta}_k 
\end{aligned}
\qquad
\begin{aligned}
I_{q\in \text{even}}^{\mu+\chi}&=
\sum_{k=2,4,\cdots}\frac{1}{2^{k-1}}
\begin{pmatrix}
k \\
\frac{k-q}{2}\\
\end{pmatrix}\bar{\chi}_k \\
I_{q\in \text{odd}}^{\eta+\nu}&=
\sum_{k=1,3,\cdots}\frac{1}{2^{k-1}}
\begin{pmatrix}
k \\
\frac{k-q}{2}\\
\end{pmatrix}\bar{\nu}_k
\end{aligned}
\end{equation}
When disorder is absent, it can be explicitly shown that all remaining \(\bar{\chi}_k\) and consequently \(I^{\mu+\chi}_{q\in\text{even}}\) vanish.
In Fig.~\ref{fig:chi} we show \(I_{q \in \text{even}}^{\mu+\chi}\) when either disorder or a uniform transverse field are present in the system (see also Fig. 2\textbf{C} of main text). Noticeably more MQC signal is presented in the disordered case than the uniform case, indicating that \(I_q^{\mu+\chi}\) can be used as a litmus test for disorder.

While the methods presented in this section are exact for extracting $D_c$($L_c$) for noninteracting systems, it works surprisingly well as an approximation for MBL systems. See Fig.~\ref{fig:sim}.\textbf{D} for comparison between exact $L_c$ and approximated $L_c$ using Eq. (\ref{eq:eqmqc}) and Eq. (\ref{eq:eqfk}). 

\section{Hamiltonian Engineering}\label{sec:AHT}
In the main paper we have used a broad range of Hamiltonians to explore different behaviors of the spin chains, ranging from integrable, to single particle and many-body localized. These Hamiltonians were obtained stroboscopically (Floquet Hamiltonians) by applying periodic rf pulse trains to the natural dipolar Hamiltonian that describes the system. We used Average Hamiltonian Theory (AHT~\cite{Haeberlen68}) as the basis for our Hamiltonian engineering method, to design the control sequences and determine the approximation errors.

To see how repeatedly applying a periodic pulse sequence modifies the dynamics of the system, we write the total Hamiltonian as \(H=H_\text{dip}+H_\text{ext}\), 
where \(H_\text{dip}=\frac{1}{2}\sum_{j<k}J_{jk}(2\sigma_z^j \sigma_z^{k}-\sigma_x^j \sigma_x^{k}-\sigma_y^j \sigma_y^{k})+\sum_j h_j \sigma_z^j\) is the system Hamiltonian, 
and \(H_\text{ext}(t)\) is the external Hamiltonian due to the rf-pulses. 
The density matrix \(\rho\) evolves under the total Hamiltonian according to \(\dot\rho=-i[H,\rho]\). 
Consider an interaction frame defined by \(\rho'={U_\text{rf}}^{\dagger}\rho U_\text{rf}\), where \(U_\text{rf}(t)=\mathcal{T}\exp[-i\int_0^t H_\text{ext}(t') dt']\) and \(\mathcal{T}\) is the time ordering operator. 
In this \textit{toggling} frame, \(\rho'\) evolves according to \(\dot{\rho}'=-i[H',\rho']\), where \(H'={U_\text{rf}}^{\dagger}H_\text{dip} U_\text{rf}\). 
Since \(U_\text{rf}\) is periodic, \(H'\) is also periodic with the same period \(t_c\). 
The evolution operator over one period can be written as \(U(t_c)=\exp[-i H_\text{F} t_c]\), where \(H_\text{F}\) is called the \textit{Floquet} Hamiltonian (or in the language of NMR the \textit{Average Hamiltonian}). 
Note that if the pulse sequence satisfies the condition \(U_\text{rf}(t_c)=1\), the dynamics of \(\rho\) and \(\rho'\) are identical when the system is viewed stroboscopically, i.e., at integer multiples of \(t_c\). 
The system evolves as if  under a time-independent Hamiltonian \(H_\text{F}\). 
To calculate \(H_\text{F}\) we employ the Magnus expansion as is usual in AHT: \(H_\text{F}=H_0+H_1+\cdots\). 
The first two terms are given by
\begin{align*}
H_0=\frac{1}{t_c}\int_0^{t_c}H'(t)dt ,\quad H_1=\frac{-i}{2t_c}\int_0^{t_c}dt_2\int_0^{t_2}dt_1[H'(t_2),H'(t_1)].
\end{align*}
The zeroth order of the average Hamiltonian \(H_0\) is often a good approximation to the Floquet Hamiltonian \(H_\text{F}\), as the first order can be set to zero by simple symmetrization of the pulse sequence. 

The basic building block of the pulse sequences we used in this work is given by a 4-pulse sequence~\cite{Kaur12,Yen83} originally developed to study MQC.
We denote a generic 4-pulse sequence as \(P(\tau_1,{\bf n}_1,\tau_2,{\bf n}_2,\tau_3,{\bf n}_3,\tau_4,{\bf n}_4,\tau_5)\), where \({\bf n}_j\) represents the direction of the \(j\)-th \(\pi/2\) pulse, and \(\tau_j\)'s the delays interleaving the pulses. In our experiments, the \(\pi/2\) pulses have a width \(t_w\) of typically 1 \(\mu\)s. \(\tau_j\) starts and/or ends at the midpoints of the pulses (see also Fig.~\ref{fig:mqcd}). In this notation, our 16-pulse sequence can be expressed as
\begin{gather*}
P(\tau_1,{\bf x},\tau_2,{\bf y},2\tau_3,{\bf y},\tau_2,{\bf x},\tau_4)P(\tau_4,{\bf x},\tau_2,{\bf y},2\tau_3,{\bf y},\tau_2,{\bf x},\tau_1)P(\tau_1,{\bf \bar{x}},\tau_2,{\bf \bar{y}},2\tau_3,{\bf \bar{y}},\tau_2,{\bf \bar{x}},\tau_4)P(\tau_4,{\bf \bar{x}},\tau_2,{\bf \bar{y}},2\tau_3,{\bf \bar{y}},\tau_2,{\bf \bar{x}},\tau_1)
\end{gather*}
where \(\{{\bf \bar{x}},{\bf \bar{y}}\}\equiv \{{\bf -x},{\bf -y}\}\). The delays are given by
\begin{gather*}
\tau_1=\tau(1+3g-v+w), \quad
\tau_2=\tau(1-u+v), \quad
\tau_3=\tau(1+u-w), \quad
\tau_4=\tau(1-3g-v+w)
\end{gather*}
where \(\tau\) is typically 4 \(\mu\)s. The cycle time \(t_c\), defined as the total time of the sequence, is given by \(t_c=24\tau\). \(u\), \(v\), \(w\), and \(g\) are dimensionless adjustable parameters, they are restricted such that none of the inter-pulse spacings becomes negative.

For our pulse sequence with finite pulse width, \(H_0\) is given by
\begin{align*}
H_0=\frac{1}{2}\sum_{j<k}J_{jk}\left[(u-w)\sigma_x^j \sigma_x^k+(v-u)\sigma_y^j \sigma_y^k +(w-v)\sigma_z^j \sigma_z^k\right]+g\sum_j h_j \sigma_z^j,
\end{align*}
and \(H_1=0\) (the first order can always be set to zero by a proper symmetrization of the sequence). 
Restricting to only nearest-neighbor (NN) terms and setting \(w=-v\) leads to Eq.(1) in the main text. 

A uniform transverse field can be introduced in two ways. One strategy is to simply apply pulses off-resonance, so that the resulting \(H_0\) contains the term \(-g\Delta \omega /2\sum_j \sigma_z^j\), where \(\Delta \omega\) is the resonance offset. This approach is easy to implement, but it cannot achieve independent control over the disordered and uniform fields, and it can result in lower-quality pulses. We use a second approach which is based on phase-shifting the entire pulse sequence. Consider rotating the \(n\)-th cycle of the pulse sequence by \((n-1)\phi\) around the \({\bf z}\) axis, which can be accomplished by phase shifting all the pulse directions ${\bf n}_j$ in the \(n\)-th cycle by \((n-1)\phi\). The evolution operator for each cycle is given by
\begin{align*}
U_1=e^{-i H_0 t_c}, \quad 
U_2=e^{-i\frac{\phi}{2}Z}e^{-i H_0 t_c}e^{i\frac{\phi}{2}Z}, \quad
U_3=e^{-i\phi Z}e^{-i H_0 t_c}e^{i\phi Z}, \quad \cdots  \quad
U_n =e^{-i(n-1)\frac{\phi}{2}Z}e^{-i H_0 t_c}e^{i(n-1)\frac{\phi}{2}Z}
\end{align*}
where \(Z=\sum_j\sigma_z^j\). The total evolution operator over \(n\) cycles is given by the product:
\begin{align*}
U(n t_c)&=U_n U_{n-1}\cdots U_3 U_2 U_1=e^{-i n\frac{\phi}{2}Z}\left[e^{i \frac{\phi}{2}Z} e^{-i H_0 t_c} \right]^n \approx e^{-i n\frac{\phi}{2}Z}e^{ -i \left(H_0-\frac{\phi}{2 t_c}Z\right)nT}=e^{-i n\frac{\phi}{2}Z}e^{-i H_\text{t} n t_c},
\end{align*}
where the total Hamiltonian is given by \(H_\text{t}=H+b Z\), with \(b=-\frac{\phi}{2t_c}\). The rotation approach also generates an extra term \(e^{-i n \phi Z/2}\), this term can be canceled in MQC experiments by rotating the encoding pulse by \(n\phi\). This approach allows us to independently tune the disordered field by adjusting \(g\), and the uniform field by varying \(b\).

We note that our methods can be applied more broadly to engineer desired Hamiltonians $ H_{des}$ using only collective rotations of the spins applied to the naturally occurring Hamiltonian, $H_{nat}$. The engineered Hamiltonian is obtained by piece-wise constant evolution under rotated versions of the natural Hamiltonian under the condition
\(
\sum_k R_k H_{nat} R_k^\dag = H_{des},
\)
where $R_k$ are collective rotations of all the spins, which achieves the desired operator to first order in a Magnus expansion. Symmetrization of the sequence can further cancel out the lowest order correction. 
Using only collective pulses limits which Hamiltonians  can be engineered, due to symmetries of the natural Hamiltonian and the action of collective operators. For typical two-body interactions of spin-1/2, an efficient tool to predict which Hamiltonian are accessible is to use spherical tensors~\cite{Ajoy13l}.

\bibliography{Biblio}

\end{document}